# Probing the spin dimensionality in single-layer CrSBr van der Waals heterostructures by magneto-transport measurements

*Carla Boix-Constant, Samuel Mañas-Valero,\* Alberto M. Ruiz, Andrey Rybakov, Krzysztof Aleksander Konieczny, Sébastien Pillet, José J. Baldoví, Eugenio Coronado.\**

C. Boix-Constant, Dr. S. Mañas-Valero, A. M. Ruiz, A. Rybakov, Dr. J. J. Baldoví, Prof. E Coronado
Instituto de Ciencia Molecular (ICMol), Universitat de València, Catedrático José Beltrán 2, Paterna 46980, Spain.
E-mail: samuel.manas@uv.es, eugenio.coronado@uv.es

Dr. K A. Konieczny, Dr. S. Pillet
Université de Lorraine, CNRS, CRM2, 54500 Nancy, France.

Two-dimensional (2D) magnetic materials offer unprecedented opportunities for fundamental physics and applied research in spintronics and magnonics. Beyond the pioneering studies on 2D $CrI_3$ and $Cr_2Ge_2Te_6$, this emerging field has expanded to 2D antiferromagnets exhibiting different spin anisotropies and textures. Of particular interest is the layered metamagnet CrSBr, a relatively air-stable semiconductor formed by antiferromagnetically-coupled ferromagnetic layers ($T_c$~150 K) that can be exfoliated down to the single-layer. It presents a complex magnetic behavior with a dynamic magnetic crossover leading to a low-temperature hidden-order below T\*~40 K. Here, we inspect the magneto-transport properties of CrSBr vertical heterostructures in the 2D limit. Our results demonstrate the marked low-dimensional character of the ferromagnetic monolayer, with short-range correlations above $T_c$ and an Ising-type in-plane anisotropy, being the spins spontaneously aligned along the easy-axis *b* below $T_c$. By applying moderate magnetic fields along *a* and *c* axes, a spin reorientation occurs, leading to a magnetoresistance enhancement below T\*. In multilayers, a spin-valve behavior is observed, with negative magnetoresistance strongly enhanced along the three directions below T\*. These results show that CrSBr monolayer/bilayer provides an ideal platform for studying and controlling field-induced phenomena in two-dimensions, offering new insights regarding 2D magnets and their integration into vertical spintronic devices.

# 1. Introduction

A CrSBr single-layer is formed by the following sextuple arrangement of atoms along the *c* axis: bromine/chromium/sulfur/sulfur/chromium/bromine (**Figure 1.a-b**). Every chromium is hexacoordinated and linked to the nearest chromium neighbor through sulfur and bromine along *a* (forming 95º and 89º, respectively) and just by sulfur along *b* (160º) and *c* (97º).[1] Although CrSBr was first described in 1990's by Beck[2] and their bulk magnetic properties were soon described by Göser *et al.*,[3] this layered material has recently gained much attention due to its magnetic, optical and electronic properties in the 2D limit.[4–7] The single-layer orders ferromagnetically ($T_c$~150 K).[8] However, in bulk, the layers couple between them antiferromagnetically –A type antiferromagnetism, $T_N$~140 K– behaving as a metamagnet, where it is possible to switch the magnetization of the layers from antiparallel to parallel configurations with an external magnetic field ($B_{sat,a}$~1.0 T, $B_{sat,b}$~0.6 T and $B_{sat,c}$~2.0 T at 2 K, where $B_{sat}$ states for the saturation fields; **Figure 1.d**).[9–11]

Regarding magneto-transport experiments, Telford *et al.* observed large negative magnetoresistance (MR) in bulk due to the layered antiferromagnetism exhibited by CrSBr.[11] When the system is thinned down, this negative MR is maintained up to the bilayer case, as has been recently shown by Telford *et al.*[7] and Ye *et al.*[12], but turns to be positive in the monolayer for fields applied along the *a* and *c* direction, remaining zero along *b*.[7] However, this scenario is more intriguing since, recently, Wu *et al.* reported on the existence of both positive and negative MR in CrSBr thin-layers depending on the current direction (parallel to *a* or *b* axes).[13] In all these cases, a striking enhancement of the MR below T*~30-40 K, quoted as a *hidden-order,* is observed.[7,13] Telford *et al.* ascribed it to the ferromagnetic ordering of magnetic defects[7] although Wu *et al.* consider its origin in terms of incoherently coupled 1D electronic chains.[13] In this line, López-Paz *et al.* have associated the hidden-order in bulk CrSBr to a spin-dimensionality crossover caused by a slowing down of the magnetic fluctuations and an eventually spin freezing at T*.[10] In addition, CrSBr multi-layers have been shown to be good candidates to induce spin-polarization in graphene heterostructures.[14,15] Thus, CrSBr atomically-thin layers are an attractive low-dimensional magnetic system with fundamental open questions and appealing potential in spintronic and magnonic applications.

In this work, we investigate the magneto-transport properties of monolayer, bilayer and trilayer CrSBr with the current passing along the *c* direction by fabricating vertical van der Waals heterostructures based on 2D materials, as graphene or metallic $2H$-$NbSe_2$.

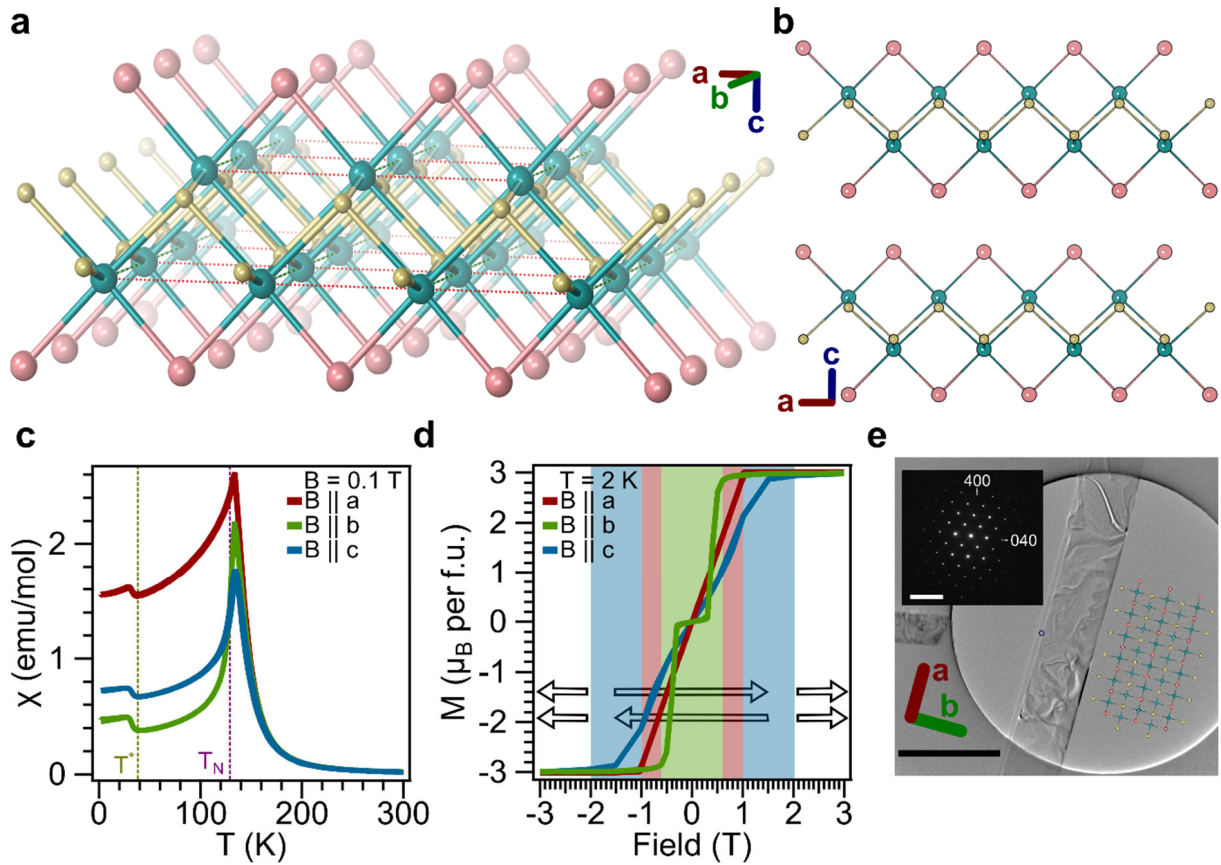

**Figure 1.- Structural and magnetic properties of bulk CrSBr.** a) Structure of a single-layer of CrSBr, where the chromium, sulfur and bromine atoms are represented as cyan, yellow and pink balls. The crystal structure of CrSBr can be described as layers of $CrS_4Br_2$ distorted octahedra, sharing S-Br edges along the *a* axis, S corners along the *b* axis, and S-S edges along the diagonal of the *ab* plane, forming a rectangular 2D magnetic lattice. b) Stacking of CrSBr layers along the *c* axis. c) Magnetic susceptibility of bulk CrSBr under an applied magnetic field of 0.1 T. Reported antiferromagnetic ($T_N$) and low-temperature hidden-order phase ($T^*$) temperatures are marked as pink and yellowish vertical dashed lines. d) Magnetization of CrSBr at 2 K. Saturation of the magnetization at 3 Bohr magnetons, corresponding to S = 3/2, is obtained when the layers are coupled ferromagnetically (represented as inset arrows). e) Transmission electron microscopy image of a CrSBr exfoliated flake together with an oriented structure cartoon of CrSBr. Scale bar: 2 µm. Inset: selected area diffraction pattern measured at the area enclosed in a blue circle. Scale bar: 10 $nm^{-1}$.

## 2. Results and Discussion

CrSBr crystals are grown by solid state techniques (see **Methods**). CrSBr crystallizes in the structure type FeOCl,[2,3,11] which is adopted by the isostructural compounds MOX (M = Ti, V, Cr, Fe; X = Cl, Br).[16–19] The whole MOX family exhibits magnetic transitions accompanied by lattice and structural distortions owing to magnetoelastic coupling.[16,17,19] We have discarded possible underlying structural phase transitions in CrSBr at low temperature by solving the crystal structure and monitoring the infrared spectra as a function of temperature (**Supplementary Section 1.2**). Next, we inspect the magnetic properties of CrSBr by incorporating them in vertical van der Waals heterostructures (**Figure 2**) based on few-layers graphene and atomically-thin metals as 2H-NbSe$_2$ or 2H-TaS$_2$ (see **Methods**).[20–22] In our experience, monolayer and bilayer CrSBr exhibits non-ohmic IV curves at low-temperatures when measured in an horizontal configuration –as expected for an intrinsic semiconductor with a bandgap of *ca*. 1.5 eV–. Thus, we employ a vertical configuration geometry, as already reported for inspecting the magnetic properties of other insulating 2D magnetic materials like CrI$_3$ or MnPS$_3$, among others.[20,23,24] All our vertical van der Waals heterostructures show linear IV curves down to 10 K (see **Supplementary Section 3**). In **Figure 2.c**, we compare the temperature dependence of MR in monolayer and bilayer CrSBr van der Waals heterostructures. Both cases exhibit almost zero MR in the paramagnetic phase for temperatures above T~240 K. Below this temperature, the MR turns negative and decreases until T$_c$~150 K is reached, when it reverts the tendency and enhances down to T~100 K. At lower temperatures, monolayer and bilayer exhibit different tendencies. In the monolayer, the MR continues its steady increase upon cooling down, being positive for fields applied along *a* and *c* and almost zero for fields along *b*. In contrast, for the bilayer, the MR remains negative and with a weak temperature variation down to T*, when it decreases considerably. This behavior does not exhibit significant differences among devices (**Supplementary Section 3.4**), observing the same trends when the current is applied along *a* or *b* (**Supplementary Section 3.3**) and independently if the van der Waals heterostructures are fabricated with few-layers graphene, 2H-NbSe$_2$ or 2H-TaS$_2$, discarding a significant influence of the substrate on the reported magneto-transport properties. We do not observe any significant CrSBr voltage-gate dependence contribution since the intrinsic graphene gate-dependence is the most significant signal (**Supplementary Section 3.4**).

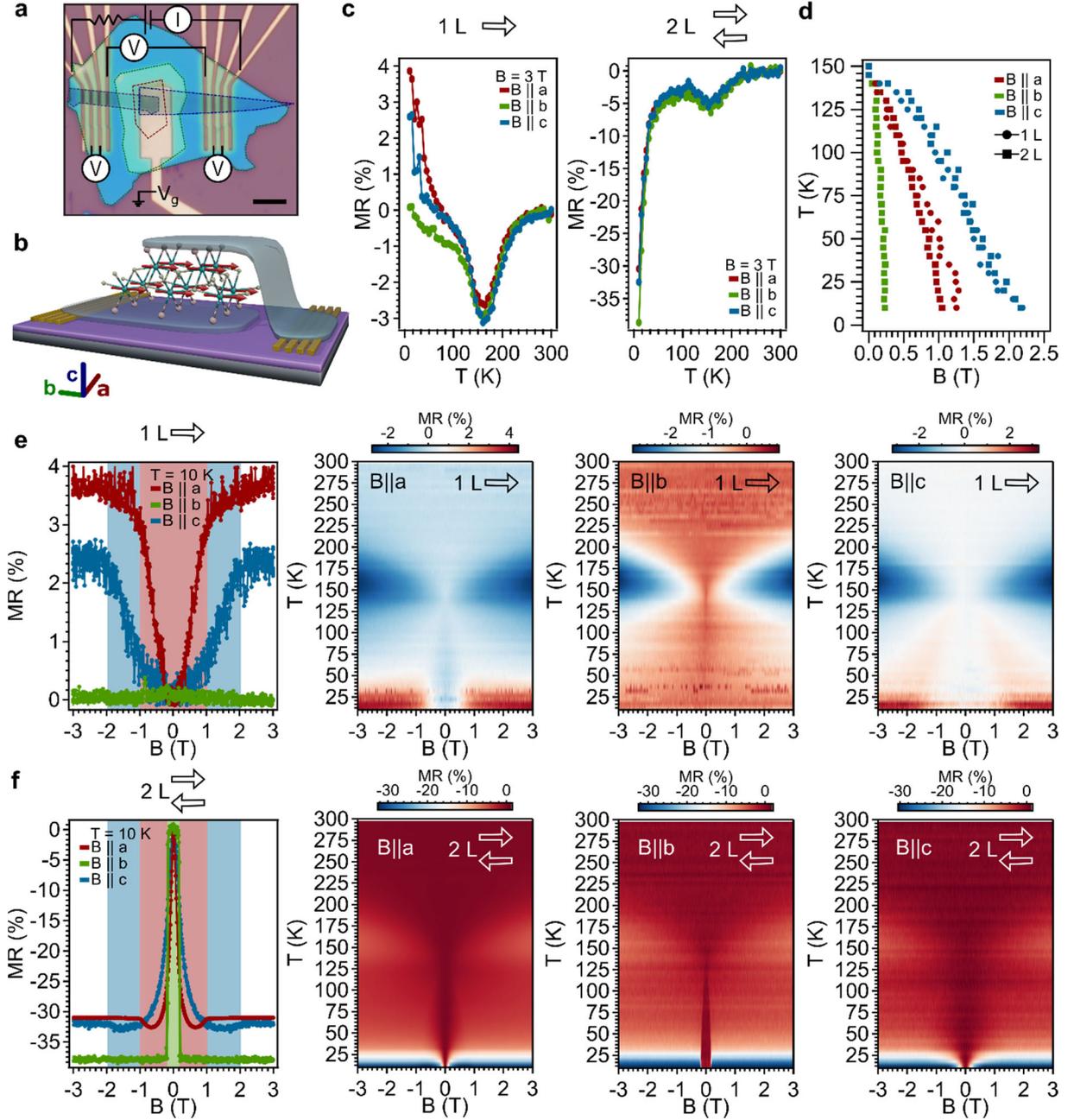

**Figure 2.- Magneto-transport properties of monolayer and bilayer CrSBr vertical van der Waals heterostructures.** a) Typical van der Waals heterostructure device. Top and bottom few-layers graphene are enclosed within blue dash lines and the CrSBr with red dashed line. The whole heterostructure is encapsulated between h-BN, marked with green dashed lines. Scale bar: 5 μm. b) Artistic view (not to scale) of the vertical van der Waals heterostructure. The few-layers graphene is represented in grey and the CrSBr structure is shown in a ball-and-stick representation (chromium, sulfur and bromine atoms are represented in cyan, yellow and pink colours, respectively). c) Temperature dependence of the magneto-resistance (MR) for a monolayer and bilayer CrSBr van der Waals heterostructure based on few-layers NbSe$_2$ (devices A.5 and B.6 in the **Supplementary Information Section 3**, respectively) for fields (3 T) applied along the different crystallographic orientations. d) Diagram for the temperature *vs.* magnetic fields values where a switching is observed in panels e and f. e-f) Field dependence of the MR at 10 K for a monolayer (device A.5) and bilayer (device B.6) CrSBr van der Waals heterostructure based on few-layers NbSe$_2$ for fields applied along the different crystallographic orientations (first panel) together with the temperature and field dependence of the MR when the magnetic field is applied along the *a* axis (second panel), the *b* axis (third panel) and the *c* axis (fourth panel). MR is defined as MR (%) = 100·[R(B) – R(0)]/R(0).

The field dependence of MR at 10 K for both monolayer and bilayer is shown in **Figure 2.e-f** (first panel). When the field is applied along *b*, the MR of the monolayer remains constant and equal to zero, within the experimental error, in agreement with the alignment of the spins along this easy-axis. This MR behavior is significantly different from those observed for the other two directions, *a* and *c*, where it keeps almost null for fields up to ~0.3 T (~0.7 T), increasing to a 4 % (2.5%) and saturating for fields above $B_{sat,a}$~1.0 T ($B_{sat,c}$~2.0 T) for *a* (*c*) direction. These saturation fields are the fields required to reorient the spins in the monolayer from *b* to *a* and *c*, respectively (**Figure 2.d**), as theoretically discussed below. In the bilayer case, the MR behavior is dominated by a metamagnetic transition where the antiferromagnetically coupled ferromagnetic layers along the out-of-plane direction undergo to a spin-flip process, *i.e.*, the system evolves from an antiferromagnetic ground state to a field-induced paramagnetic phase above a critical field, with a parallel magnetization of the two ferromagnetic layers.[25] This leads to a spin-valve behavior characterized by large MR differences (*ca.* 35%) between these two states. When the field is applied along *b*, this change in MR is abrupt and occurs at ~0.2 T. On the contrary, when the field is applied along *a* or *c*, a progressive field evolution of MR is observed, characterized by an upturn (at *ca.* 0.65 T and 1.5 T, respectively) until it saturates and shows no field evolution. These tendencies are in line with the magnetic response of bulk CrSBr (**Figure 1.d**). Notably, the saturation fields for the *a* and *c* axis are comparable for monolayer, bilayer and bulk ($B_{sat,a}$~1.0 T, $B_{sat,c}$~2.0 T). In turn, the critical field along *b* required to switch the layers, is significantly reduced from the bulk to the bilayer (from $B_{sat,b}$~0.6 T to ~0.2 T). As well, the absolute MR value is much larger for the multilayer case, exhibiting MR values higher than those reported for bulk and atomically-thin layers measured in a horizontal configuration.[7,11,13]. This can be understood within a conventional spin-valve picture, taking into account a two-channel current model: for a parallel configuration in the magnetization of the layers, the resistance across the heterostructure is smaller than for the antiparallel case.

Next, we discuss both the temperature and field dependence of MR (**Figure 2.e-f**, 2D plots). The MR is zero at room temperature, developing a negative value in the range 150 K-240 K, *i.e.*, just above the $T_c$ of the layers, which appears as wings that persist down to *ca.* 100 K. This trend is observed for all the field orientations and can be related with the appearance of short-range ferromagnetic correlations above $T_c$, underlying the low-dimensional character of these magnetic materials, being also in agreement with recent muon spin resonance and magnetic susceptibility experiments on bulk CrSBr.[10] Interestingly, MR is zero in the monolayer when the field is applied along *b* below *ca.* 100 K and does not change with temperature. This demonstrates that the spin is locked along *b* below this temperature, in agreement with bulk results.[10] In contrast, when the field is applied along *a* or *c* directions one observes positive MR above certain field threshold, which shows a temperature dependence (**Figure 2.d**). Below a temperature close to T*, this field-induced reorientation is concomitant with a small MR enhancement (*ca.* 2-4%), thus suggesting the appearance of a novel phase in which the spins of the layer are fully oriented along these two directions, while they fluctuate at higher temperatures.[10] In the bilayer case, the field-induced metamagnetic transition (from ferromagnetic to paramagnetic with a parallel alignment of the magnetization of the layers) is observed below $T_c$ with an MR enhancement below T* in the three directions (from *ca.* 4% to *ca.* 35%) above the critical fields. Interestingly, and in contrast with the monolayer, the enhancement of MR below T* is also observed in the easy-axis *b*. We speculate that this difference is due to a spin freezing of the two ferromagnetic layers in the paramagnetic phase below T*, although other alternative scenarios, as the magnetic ordering of electronic point defects,[7] are possible. Thus, in the antiferromagnetic state (below $T_N$ and below the critical field), the bilayer can be seen as formed by one layer magnetized along +*b* and the other along -*b*. When a magnetic field along +*b* is applied, the magnetization of the former layer will remain along +*b*, with the spins locked in this direction. In turn, the latter layer will be switched to +*b* above the critical field, undergoing thermal fluctuations above T*. Below T* the spins of this layer are also locked along +*b*, thus facilitating the spin transport across the bilayer. It must be noted that the enhanced MR is already present in the monolayer case but only when the system is driven out of the ground state (spins aligned along *b*) by the external magnetic field. For fields applied along *b* or fields along *a* and *c* lower than the field needed to reorient the spins, the MR is null. Thus, this interpretation differs from the spin-polarization of defects proposed by Telford *et al.*,[7] pointing towards a field-induced phenomenon.

Finally, we rotate the magnetic field along the *ab* plane with the aim of studying the magnetic anisotropy discussed above (**Figure 3**). For the monolayer, two clear regions are observed with MR~0 (blue) and MR>0 (red), for spins aligned along *b* and *a*, respectively. A different scenario is offered in the multilayer case (**Figure 3.b-c**). Both bilayer and trilayer exhibit similar trends with two-clear regions of high/low absolute MR value, that are related to the relative spin orientation between the layers (antiparallel/parallel). Upon rotation, the field-induced switching from the antiparallel to the parallel configuration occurs at slightly different magnetic fields, being it maximum/minimum when the field is aligned along the *a*/*b* axis, in agreement with the observations above. For fields below 0.2 T, there is a progressive evolution of the MR from the *a* to the *b* axis, whereas, for fields in between ~0.2 T and ~0.6 T, a sharp transition to the high MR state is observed at different rotation values, exhibiting a hysteretic behavior that depends on the value of the applied external magnetic field. Above ~1.0 T, there is only a very weak dependence of the MR upon rotation.

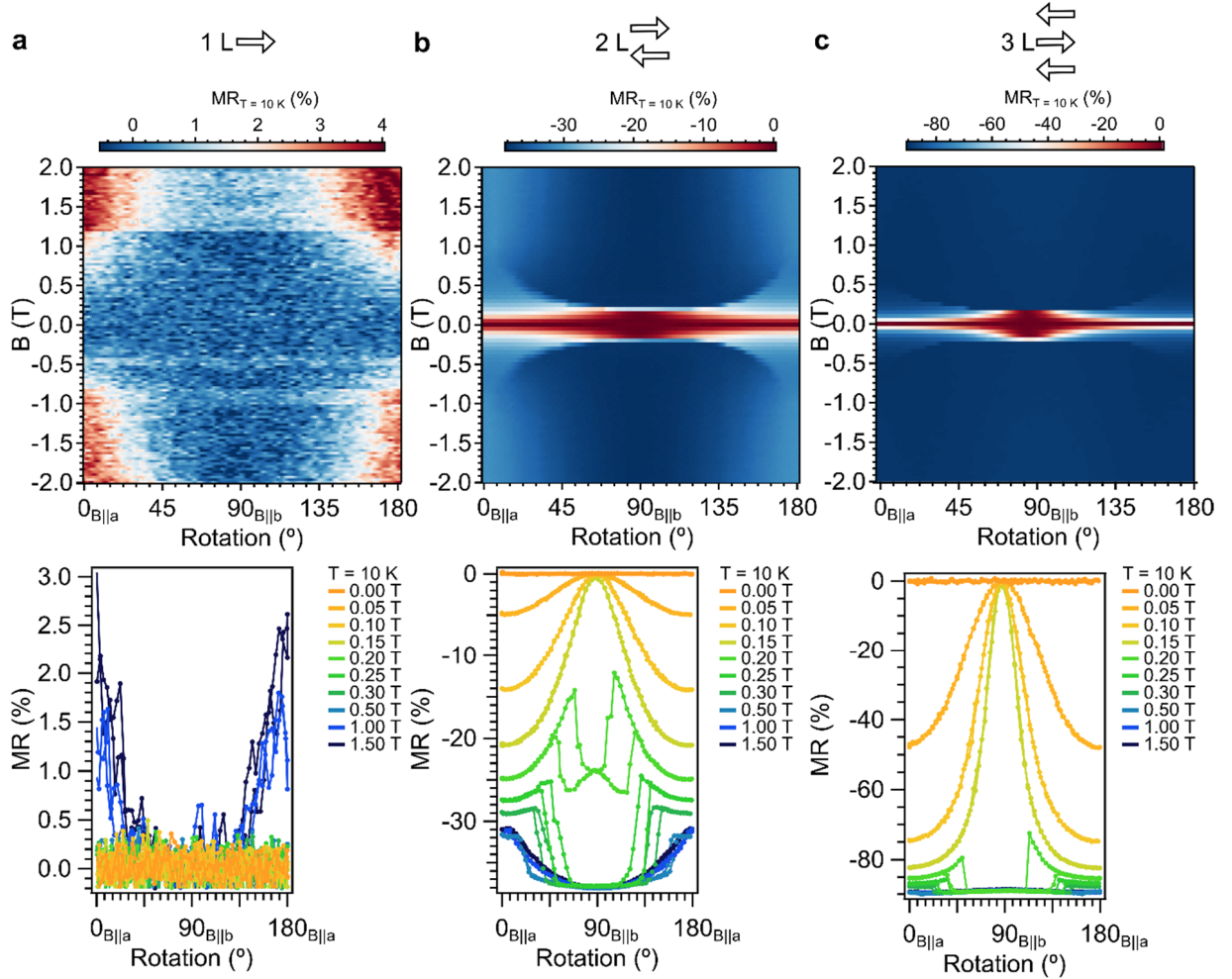

**Figure 3.- Field orientation dependence of the magneto-transport properties of CrSBr van der Waals heterostructures in the *ab* plane at 10 K.** a) Monolayer CrSBr (device A.5 in the **Supplementary Information**). b) Bilayer CrSBr (device B.6 in the **Supplementary Information**). c) Trilayer CrSBr (device C.1 in the **Supplementary Information**). Magneto-resistance (MR) is defined as MR (%) = 100·[R(B) – R(0)]/R(0).

To rationalize the field-induced spin reorientation in the CrSBr monolayer, we compute the magnetic anisotropy along the different axis by performing DFT+U calculations (see **Methods**).[1,26] These support that the most stable configuration is found for spins aligned along the *b* axis, followed by spins along *a* and *c* (anisotropy energies of 0.06 and 0.14 meV/Cr atom, respectively). Thus, the estimated magnetic field needed for reorienting the spin from the easy-axis (*b*) to the intermediate-axis (*a*) is 0.3 T and 0.8 T to the hard-axis (*c*). These values are compatible with our experimental observations. In addition, we consider the thermal evolution of the exchange network according to a Heisenberg exchange Hamiltonian with three exchange magnetic parameters (**Figure 4.a-b**). These exchange parameters are ferromagnetic (in the range 2-4 meV; **Supplementary Section 4**) and increase upon cooling down, tending to saturate at low temperatures (**Figure 4.c**). This trend mimics the evolution of the unit-cell parameters,[10] suggesting that the reported unconventional expansion along *a* by López-Paz[10] is triggered by magnetoelastic coupling (**Supplementary Section 4**).

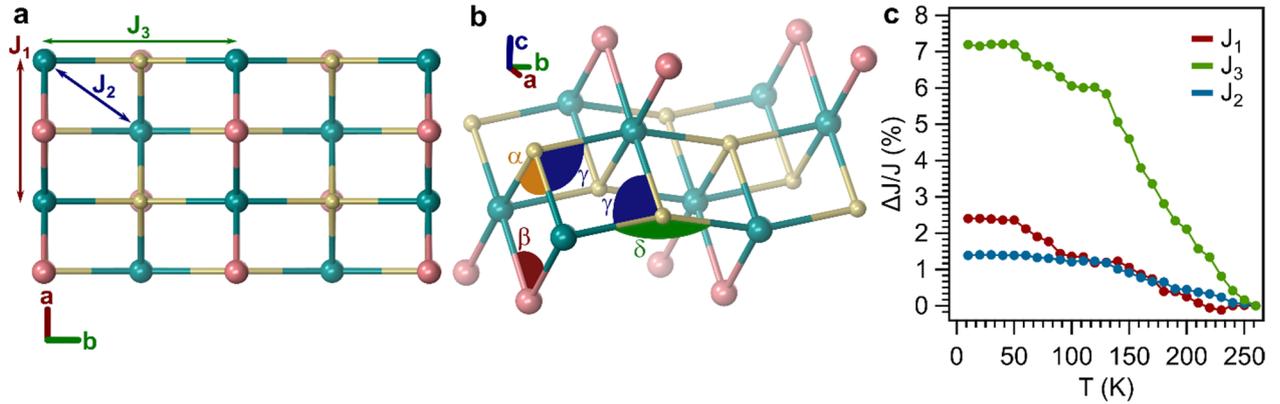

**Figure 4.- Magnetic interactions in CrSBr.** a) Top view of CrSBr monolayer pointing the $J_1$, $J_2$ and $J_3$ directions. $J_1$ is the first-neighbors exchange along *a* and it accounts for the interactions between Cr-Br-Cr (β) and Cr-S-Cr (α), while $J_2$ defines coupling between Cr atoms from different "sublayers" coupled along c (γ), and $J_3$ between Cr atoms mediated by S along the b axis (δ). b) Schematic representation of angles mediating the four different super-exchange paths. c) Change in $J_1$, $J_2$ and $J_3$ over temperature for a Hubbard U = 4 eV.

## 3. Conclusion

To sum up, we have investigated the magneto-transport properties of CrSBr vertical van der Waals heterostructures, unveiling open questions regarding the rich 2D physics exhibited by these low-dimensional materials. In particular, MR measurements in the monolayer demonstrate the moderate magnetic anisotropy of the ferromagnetic layer, with a ground state in which the spins are locked along the *b* axis below $T_c$ (Ising-type spin anisotropy), but that can be reoriented towards *a* and *c* upon application of moderate magnetic fields along these directions (~0.3 T and ~0.7 T, respectively), as supported by first-principles calculations. In the multi-layer case, a spin-valve behavior is observed. Both in the monolayer and multilayer case, an enhanced MR is observed below T* when the applied magnetic fields are larger than those required for inducing a spin-reorientation (monolayer case) or a spin-switching (bilayer case). Above T*, the reoriented spins are fluctuating due to the competition between magnetic and thermal energies whereas, below T*, the field applied along the three crystallographic axes stabilize a cooperative state in which the spins are fully oriented –frozen– along these axes, both within and between the layers. This scenario is reminiscent to that proposed by López-Paz *et al.* to explain the intriguing anomaly observed in bulk at T*,[10] but its origin is different. Here, it is a field-induced phase that depends on the magnetic anisotropy within the layer and on the interlayer magnetic interactions. Thus, this phase appears only when the spin orientation is taken out of its ground state by the application of a magnetic field. In addition, these trends are in stark contrast to what occurs in other 2D magnets with strong out-of-plane magnetic anisotropy, as CrI$_3$.[23] More generally, CrSBr magnetic anisotropy can allow different spin-textures, which is of great importance in the field of van der Waals heterostructures and proximity effects in 2D materials when combined with other 2D strongly correlated materials, as superconductors or topological insulators, or even when CrSBr layers are twisted between them. This makes CrSBr of high interest not only as a new 2D magnetic model but also as a potential spintronic component. In fact, as compared with the other known 2D magnets, this material presents several advantages: it exhibits larger chemical stability, easier exfoliation down to the monolayer and smaller switching fields in the bilayer (0.2 T in the *b* axis, compared to 0.65 T in CrI$_3$), allowing in addition to easily tune these switching fields by applying an external magnetic field along different crystal orientations.

## 4. Experimental Section

*Crystal growth:* High quality crystals of CrSBr were grown by solid state techniques, as previously reported.[11] The crystal structure was verified by powder and single crystal X-ray diffraction as well as selected area electron diffraction in atomically-thin layers together with the elemental composition by energy-dispersive X-ray spectroscopy (see **Supplementary Section 1.1** for further details). Crystals of 2H-NbSe$_2$ and 2H-TaS$_2$ were grown by chemical vapor transport (CVT) using iodine as a transport agent, as already reported by some of us.[27]

*Bulk magnetic measurements:* Variable-temperature (2–300 K) direct current (d.c.) magnetic susceptibility measurements were carried out in an applied field of 1.0 kOe, and variable field magnetization measurements up to ±5 T, at 2.0 K. All the measurements were performed with a SQUID magnetometer (Quantum Design MPMS-XL-5).

*X-ray diffraction:* Single crystal X-ray diffraction measurements were performed as a function of temperature in the 10-75K range using a Microfocus Supernova diffractometer (Mo Kα radiation, λ = 0.71073 Å) equipped with a two dimensional ATLAS detector, and a Helijet He open flow cryosystem. A needle shape single crystal sample was cut from a very large block shape single crystal. Unit cell parameters were determined from 10K to 75K (5K step), while complete data collection for structural determination and analysis was performed at 10K and 75K. Analytical absorption corrections were applied. Initial structural models were based on those proposed by Beck,[2] and refined using full matrix least squares on F2 using OLEX.[28] More experimental and refinement details are provided in the supplementary information. The corresponding CIF files can be retrieved from Cambridge Structural Database (CSD) (deposition numbers: CCDC 2161029 & 2161030). Our results are in agreement with those reported by Telford *et al.*[11]

*IR spectroscopy:* All IR measurements were performed using a Nicolet 5700 FT-IR spectrometer equipped with a He closed-cycle cryostat. The sample was grinded, mixed with polyethylene powder, pressed into pellets, and glued to the cold-finger of the cryostat using silver-paste thermal adhesive. Measurements were carried out as a function of temperature from 300K to 10K in the 50-1000 cm-1 wavenumber range.

*van der Waals heterostructure fabrication:* 2D layers are obtained by mechanical exfoliation from their bulk counterparts under strict inert conditions since CrSBr atomically-thin layers degrade in air.[7] The obtained flakes were examined by optical microscopy (NIKON Eclipse LV-100 optical microscope under normal incidence) as a fast tool for identifying the number of layers. The optical contrast was calculated as $C = (I_{flake} - I_{substrate})/(I_{flake} + I_{substrate})$,[22] with $I_{flake}$ being the intensity of the 2D material and $I_{substrate}$ the intensity of the substrate. The maximum optical contrast for a CrSBr monolayers is observed upon 550 nm illumination (see **Supplementary Section 2** for further details). Atomic force microscopy images were taken with a Nano-Observer AFM from CSI Instruments. Typical CrSBr flakes exhibit a ribbon shape, being the long (short) direction associated with the a (b) axis, as verified by optical contrast, Raman spectroscopy and selected area electron diffraction patterns obtained by transmission electron microscopy (**Figure 1.e**, **Supplementary Section 1.3** and **Supplementary Section 2**). The van der Waals heterostructures are fabricated by assembling the different layers by the deterministic assembly of the flakes using polycarbonate, as reported by Wang *et al.*,[29] using a micromanipulator. Finally, the stack of 2D materials (for example, h-BN/few-layers graphene/CrSBr/few-layers graphene) was placed on top of pre-lithographed electrodes (5 nm Ti/50 nm Au on 285 nm $SiO_2$/Si from NOVA Electronic Materials, LCC). The whole process was performed under inert atmosphere conditions (argon glovebox). Importantly, the heterostructures are encapsulated with hexagonal boron-nitride thin-layers for avoiding any possible degradation.

*TEM:* Mechanical exfoliated flakes were transferred onto a silicon nitride (50 nm thick) membrane, as explained above. TEM images and diffraction patterns were acquired with a JEOL JEM-2100F with a field-emission gun operating at 200 kV. Simulated SAED patterns were generated with SingleCrystal software.

*Electrical measurement setup:* Electrical measurements were performed in a Quantum Design PPMS-9 cryostat with a 4-probe geometry, where a DC current was passed by the outer leads and the DC voltage drop was measured in the inner ones. DC voltages and DC currents were measured (MFLI from Zurich Instruments) using an external resistance of 1 MΩ, i.e., a resistance much larger than the sample. Temperature sweeps were performed at 1 K·min$^{-1}$, field sweeps at 200 Oe/s and rotation sweeps at 3 °/s. All our van der Waals heterostructure devices (14 devices in total) exhibit ohmic behavior from room to low temperatures (**Supplementary Section 3.4**), in contrast to the sigmoidal IV curves reported for horizontal CrSBr devices.[13] We note that, for the van der Waals heterostructures based on few-layers graphene, the intrinsic graphene magneto-resistance features can be observed, especially for fields applied along the *c* direction. In addition, in the monolayer case, we observe a narrow peak centered at zero field for the few-layers graphene-based van der Waals heterostructures, that is absent in the metallic-based heterostructures incorporating 2H-NbSe$_2$ or 2H-TaS$_2$ instead of graphene. This can be attributed to a spin-polarized current in the few-layers graphene due to proximity effect with the CrSBr, as reported by Ghiasi *et al*.[7]

For the resistance data, we consider different transport models (as energy-activated Arrhenius law or hopping mechanisms; see **Supplementary Section 3.5**), where different slope changes can be related with the magnetic ordering temperatures. Overall MR trends are comparable with previous measurements in horizontal devices.[7] Nonetheless, raw resistance curves are different to the ones reported for atomically-thin layers measured in horizontal geometry (see comparative in **Supplementary Section 3.2**). This supports sensing the out-of-plane component (*c* axis), as already reported for other vertical van der Waals heterostructures.[20,21,23,30,31]

In some bilayer devices, we observe certain hysteresis along the *b* axis (**Supplementary Section 3**). The width of the hysteresis varies slightly between devices (from 0 T to ~0.07 T) and can be attributed to a non-perfect alignment between the magnetic field and the crystal axis, as corroborated by rotation experiments (**Figure 3**). MR is defined as: MR = 100[R(B) – R(0)]/R(0), where B is the external magnetic field.

*Computational details:* We carried out first principles calculations based on spin-polarized density functional theory (DFT) in the plane wave formalism as implemented in the QuantumESPRESSO package.[32] The exchange-correlation energy is described by the generalized gradient approximation (GGA) using the Perdew–Burke–Ernzerhof (PBE)[33] functional and standard ultra-soft (USPP) solid-state pseudopotentials. The important role played by the strong correlations of the *d* electrons of the $Cr^{3+}$ magnetic ions is simulated with a Hubbard-corrected DFT+U approach, choosing a standard on-site Hubbard U of 4 eV in the simplified version proposed by Dudarev *et al.*[34] We also evaluated U = 3 and 5 eV (see **Supplementary Section 4**). The electronic wave functions were expanded with well-converged kinetic energy cut-offs for the wave functions (charge density) of 60 (600). Given the DFT limitation of working at 0 K, we approximate the effect of temperature to the gradual variation of parameters, extrapolating these data directly from experimental measurements and optimizing the atomic coordinates.[10] We used the experimental bulk lattice parameters determined by López-Paz *et al.*[10] between 10 K and 270 K and optimized the atomic positions using the Broyden-Fletcher-Goldfarb-Shanno (BFGS) algorithm until the forces on each atom were smaller than $1 \cdot 10^{-5}$ Ry/au and the energy difference between two consecutive relaxation steps was less than $1 \cdot 10^{-7}$ Ry. To avoid unphysical interactions between images along the non-periodic direction, we add a vacuum of 18 Å in the *z* direction. The Brillouin zone was sampled by a fine Γ-centered 8 × 8 × 1 k-point Monkhorst–Pack.[35] In order to determine the exchange parameters, we performed DFT+U+SOC calculations using fully-relativistic ultrasoft pseudopotentials with the spin oriented in each of the three spatial directions. For each one, we constructed a tight-binding model based on maximally localized Wannier functions[36] using the d orbitals of Cr, and p orbitals of S and Br as implemented in the Wannier90 package.[37] This allowed us to determine the isotropic Heisenberg parameters using the Green's function method as implemented in TB2J package[38] within a 20x20x1 supercell. Magnetic anisotropy energy (MAE) was obtained by performing self-consistent calculations with SOC and norm-conserving pseudopotentials, which were extracted from the Pseudo-Dojo database.[39] Then, we evaluated the energy differences between the states with spins aligned along the three main crystallographic axes. For ensuring accurate MAE convergence (~0.02 meV/Cr) we set up a kinetic energy cutoff of 190 Ry and a Monkhorst-Pack grid of 30x30x1 k-mesh.

**Supporting Information**

Supporting Information is available from the author.

**Acknowledgements**

The authors acknowledge the financial support from the European Union (ERC AdG Mol-2D 788222, FET OPEN SINFONIA 964396, ERC-2021-StG-1010426820 2D-SMARTiE), the Spanish MICINN (2D-HETEROS PID2020-117152RB-100, co-financed by FEDER, and

Excellence Unit "María de Maeztu" CEX2019-000919-M), the Generalitat Valenciana (PROMETEO Program and PO FEDER Program, IDIFEDER/2018/061, grant CDEIGENT/2019/022 and pre-doctoral fellowship GRISOLIAP/2021/038), the CNRS, the "Agence Nationale de la Recherche" (Project MolCoSM : Grant ANR-20-CE07-0028-03), and the Region Grand-Est. C.B.-C. thanks the Generalitat Valenciana for a Ph.D fellowship.

We thank J. M. Martínez-Agudo, G. Agustí and Á. López-Muñoz for their constant technical support and helpful discussions as well as S. Ferrer-Nicomedes regarding the optical contrast calibration of CrSBr atomically-thin layers and device fabrication.